\documentclass[9pt, conference]{IEEEtran}
\IEEEoverridecommandlockouts

\usepackage{cite}
\usepackage{amsmath,amssymb,amsfonts}
\usepackage{algorithmic}
\usepackage{graphicx}
\usepackage{textcomp}
\usepackage{xcolor}
\usepackage{hyperref}
\usepackage{booktabs} 
\usepackage{enumitem}
\usepackage{graphicx}
\usepackage{bm}
\def\BibTeX{{\rm B\kern-.05em{\sc i\kern-.025em b}\kern-.08em
    T\kern-.1667em\lower.7ex\hbox{E}\kern-.125emX}}

\newcommand{\mypar}[1]{\noindent{\bf #1}}

\makeatother

\begin{document}

\title{Audio Texture Manipulation by Exemplar-Based Analogy}

\author{
\IEEEauthorblockN{Kan Jen Cheng\textsuperscript{*},
Tingle Li\textsuperscript{*},
Gopala Anumanchipalli}
\IEEEauthorblockA{
University of California, Berkeley\\
\small{\tt \url{https://berkeley-speech-group.github.io/audio-texture-analogy/}}
\thanks{\textsuperscript{*} Equal Contribution}
}}

\maketitle

\begin{abstract}
Audio texture manipulation involves modifying the perceptual characteristics of a sound to achieve specific transformations, such as adding, removing, or replacing auditory elements. In this paper, we propose an exemplar-based analogy model for audio texture manipulation. Instead of conditioning on text-based instructions, our method uses paired speech examples, where one clip represents the original sound and another illustrates the desired transformation. The model learns to apply the same transformation to new input, allowing for the manipulation of sound textures. We construct a quadruplet dataset representing various editing tasks, and train a latent diffusion model in a self-supervised manner. We show through quantitative evaluations and perceptual studies that our model outperforms text-conditioned baselines and generalizes to real-world, out-of-distribution, and non-speech scenarios.
\end{abstract}

\begin{IEEEkeywords}
Audio Texture Manipulation and Generation, Exemplar-Based Analogy, Self-Supervised Learning.
\end{IEEEkeywords}

\section{Introduction}
Audio texture manipulation \cite{mcdermott2011sound, sharma2022trends} refers to the process of altering the overall perceptual quality of a sound, shaped by
the number and interaction of different sound sources, to achieve a desired outcome, such as enhancing clarity, removing specific elements, or modifying the sound’s content. This capability has a broad range of applications, including sound design and editing. Consider a scenario where an audio recording contains two distinct bird sounds: a sparrow and a crow. The goal might be to remove the crow's sound while preserving the sparrow's, but existing audio editing methods, particularly text-conditioned models~\cite{jin2017voco, wang2023audit}, struggle to perform this task. This is because users may not be skilled at providing precise prompts, but rather they may give vague instructions like ``remove the bird sound," which leads to the removal of all bird sounds instead of just the crow's. Furthermore, the heavy reliance on human-labeled data can introduce errors \cite{demartini2021managing}, as these annotations are often subjective. For example, the widely used BBC SFX dataset~\cite{bbc2017} contains bird-labeled entries that do not actually include any bird sounds, potentially misleading the model.

A promising alternative to text-conditioned approaches is exemplar-based analogy \cite{skousen2012analogical}. This classic framework has been widely explored in vision, where transformations are applied to new data by comparing pairs of before-and-after examples. Tasks like image analogy~\cite{hertzmann2001image}, inpainting~\cite{bar2022visual}, and visual sentences~\cite{bai2024sequential} have demonstrated how models can generalize complex transformations by learning from paired examples. In audio, similar exemplar-based methods have been employed for conditional Foley synthesis~\cite{du2023conditional} and soundscape stylization~\cite{li2024self}, where the objective is to transfer certain auditory properties from one sound to another. Exemplar-based models allow for more intuitive and flexible control over transformations compared to text-based methods, as they avoid the ambiguity and subjectivity of human language by using clear before-and-after examples. Moreover, exemplar analogy has recently shown promise in prompting and guiding large language models by offering chain-of-thought reasoning~\cite{wei2022chain, zhang2022automatic, yasunaga2023large}, further highlighting the potential for exemplar-based methods to handle complex transformations in various domains.

\begin{figure}[t]
\centering  
\includegraphics[width=\linewidth]{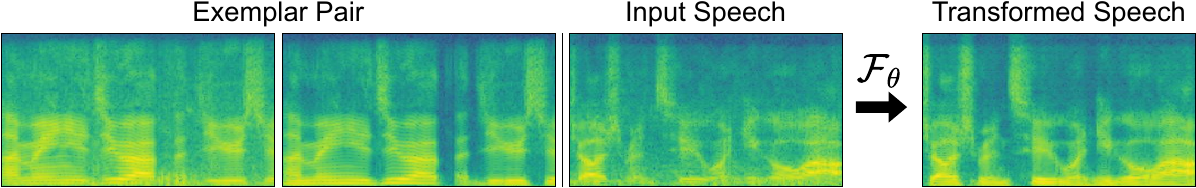}  
\caption{{\bf Exemplar-based analogy for audio texture manipulation.} We manipulate input speech (middle) based on an exemplar pair (left), where the pair defines the desired transformation such as adding, removing, or replacing specific sound elements.}  
\label{fig:analogy}
\vspace{-2mm}
\label{fig}
\end{figure}

Inspired by these successes, we propose a model to audio texture manipulation via exemplar-based analogy. Our method leverages audio examples, rather than text instructions, to define the desired transformation. Specifically, we provide a pair of exemplar audio clips: the first representing the original sound (pre-editing), and the second illustrating the desired result (post-editing). The model then learns to apply the same type of transformation to new input audio. Whether the task involves removing, adding, or modifying specific sound textures, our model enables precise and controlled manipulation by example. To achieve this, we construct a quadruplet dataset combining speech from LibriSpeech \cite{panayotov2015librispeech} and VCTK \cite{yamagishi2019vctk} with ambient textures from BBC SFX, where each sample includes an exemplar input, an exemplar output, a new input audio, and the transformed output. These quadruplets cover three common editing tasks and are used to train a latent diffusion model~\cite{rombach2022high} in a self-supervised manner, where the exemplar pair acts as a guide for the transformation (Fig.~\ref{fig:analogy}). This setup allows the model to learn how to infer the transformation from the exemplar pair and apply it to new inputs.

After training, we assess our model’s performance by manipulating the audio texture in response to the exemplar pair. Through both quantitative evaluations and perceptual studies, we show that our model:
\begin{itemize}[topsep=0pt, noitemsep, leftmargin=26pt]
\item Learns to successfully manipulate audio textures from exemplar pairs.
\item Achieves comparable or better performance to text-conditioned models across different editing tasks.
\item Generalizes to out-of-distribution, real-world, and non-speech scenarios.
\end{itemize}

\section{Exemplar-Based Audio Texture Manipulation}
Our goal is to manipulate an input audio based on transformations defined by a pair of exemplar audios, representing the ``before" and ``after" states of the desired transformation. We learn a function $\mathcal{F}_\theta(\bm{a}_{q}, \bm{a}_{e}^1, \bm{a}_{e}^2)$, parameterized by $\theta$, that modifies an input audio $\bm{a}_{q}$ based on an exemplar pair $(\bm{a}_{e}^1, \bm{a}_{e}^2)$, which defines the transformation direction (e.g., adding, removing, or replacing specific sound elements). We show that $\mathcal{F}_\theta$ can be learned solely from unlabeled audios.

\begin{figure}[t]
\centering  
\includegraphics[width=\linewidth]{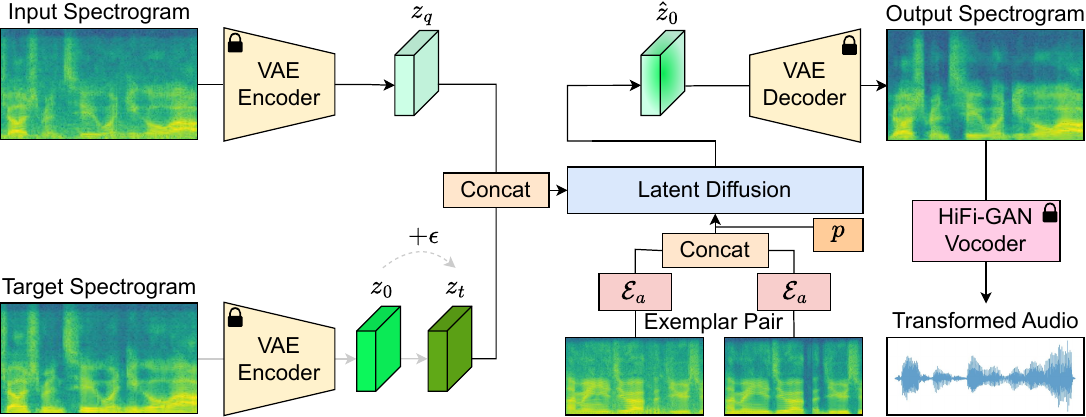}  
\caption{{\bf Model architecture.} Given the input audio and exemplar pair, our goal is to transform the input to match the texture transformation demonstrated by the exemplar pair. We employ a pre-trained VAE encoder to encode both the input and target spectrograms to the latent space, and feed them into a latent diffusion model together with the exemplar pair embedding and positional encoding. Finally, we use pre-trained VAE decoder and HiFi-GAN vocoder to reconstruct the waveform from the latent space. Note that the VAE encoder for the target spectrogram is not used at test time.}  
\label{fig:model}  
\vspace{-2mm}
\end{figure}

\subsection{Self-Supervised Audio Texture Manipulation}
We propose a self-supervised pretext task that trains a model to manipulate audio textures by learning from exemplar pairs that specify transformations. The model learns to infer the transformation direction from the exemplar pair and apply it to new input audio.

To achieve this, we first construct quadruplets consisting of: an exemplar input audio $\bm{a}_{e}^1$, an exemplar output audio $\bm{a}_{e}^2$, an input audio $\bm{a}_{q}$ to be transformed, and the corresponding target audio $\bm{a}_{o}$, which represents the transformed version of the input audio. The exemplar pair $(\bm{a}_{e}^1, \bm{a}_{e}^2)$ defines the specific transformation to be applied, such as adding, removing, or replacing auditory elements.

Through this self-supervised task, we empirically find that the model learns to tailor its output according to the transformation implied by the exemplar pair, which aligns with the assumption that the exemplar pair is informative about the input. This finding allows us to substitute in a completely {\em different} exemplar pair at test time.

\subsection{Exemplar-Based Manipulation Model}
Our model $\mathcal{F}_\theta$ consists of three main components: \romannumeral1) encoding the input and exemplar audios into a latent space; \romannumeral2) manipulating the input audio’s latent representation using a latent diffusion model conditioned on the exemplar pair; \romannumeral3) reconstructing the transformed audio from the latent space. We train this model to align the latent representations of the transformed audio and the desired output, conditioned on the transformation defined by the exemplar pair.

\mypar{Conditional latent diffusion model.}
We train a conditional diffusion model to manipulate audio texture in the latent space. Similar to prior work in denoising diffusion probabilistic model \cite{ho2020denoising} and latent diffusion models~\cite{rombach2022high}, our model operates directly on the encoded latent of mel-spectrograms, which enables efficient transformation of the sound texture.

The model takes as input the encoded latent of the target audio, $\bm{z}_{0} = \text{Enc}(\bm{a}_{o})$, along with the exemplar pair $(\bm{a}_{e}^1, \bm{a}_{e}^2)$, which determines the direction of the transformation. A random denoising step $t$ and random noise $\bm{\epsilon} \sim \mathcal{N}(\mathbf{0}, \mathbf{I})$ are applied through a noise schedule \cite{song2020denoising}, producing a noisy latent $\bm{z}t$. The model is then tasked with predicting the noise $\bm{\epsilon}$ that was added to the noisy latent $\bm{z}_t$, guided by the input audio $\bm{a}_q$ and the exemplar pair $(\bm{a}_{e}^1, \bm{a}_{e}^2)$, using the following loss function $\mathcal{L}_\theta$:

\begin{equation}
\label{eq:loss_diffusion}
    \mathcal{L}_\theta = \mathbb{E}_{\bm{z}_0,\bm{a}_e^1, \bm{a}_e^2, \bm{\epsilon} \sim \mathcal{N}(\mathbf{0}, \mathbf{I}), t}\Vert\bm{\epsilon} - \bm{\epsilon}_\theta(\bm{z}_t, t, \bm{z}_{q},\bm{a}_e^1, \bm{a}_e^2)\Vert^{2}_{2}
\end{equation}

\mypar{Mel-spectrogram compression.}
We utilize a ResNet-based \cite{he2016deep} VAE \cite{kingma2013auto} to compress the mel-spectrogram \cite{stevens1937scale} $\bm{a} \in \mathbb{R}^{T\times F}$ into a latent space $\bm{z}\in\mathbb{R}^{{T/r}\times{F/r}\times d}$, where $r$ is the compression level, $T/r$ and $F/r$ represents a lower-resolution time-frequency bin, and $d$ denotes the embedding size at each bin. In our experiments, we employ a pre-trained VAE model from AudioLDM \cite{liu2023audioldm}.

\mypar{Representing exemplar pair.}
We use CLAP audio encoders \cite{elizalde2023clap} $\mathcal{E}_a(\cdot)$ to extract exemplar audio embeddings $\mathcal{E}_a(\bm{a}_e^1) \in \mathbb{R}^L$ and $\mathcal{E}_a(\bm{a}_e^2) \in \mathbb{R}^L$, where the embedding size is $L$. Each embedding is then applied linear projection, concatenated, and input into the diffusion model via the cross-attention mechanism \cite{vaswani2017attention}.

\mypar{Learnable positional encoding.}
In order to encourage the model to be aware of the temporal relationship between the exemplar pair, we introduce a learnable positional encoding $\bm{p}$ \cite{devlin2018bert} to signify which audio corresponds to the ``before" (pre-editing) state and which corresponds to the ``after" (post-editing) state. This encoding is necessary for guiding the model's understanding of the transformation direction, allowing it to apply the changes to the input audio.

\mypar{Classifier-free guidance.}
We apply classifier-free guidance \cite{ho2022classifier} to ensure the generated audios attain a trade-off between diversity and quality. During training, we randomly disable the condition with a probability of 10\%. At test time, we adjust the score estimates according to a guidance scale ($\lambda \geq 1$), leading them towards the conditional $\bm{\epsilon}_\theta(\bm{z}_t, t, \bm{z}_{q}, \bm{a}_e^1, \bm{a}_e^2)$ and away from the unconditional $\bm{\epsilon}_\theta(\bm{z}_t, t, \bm{z}_{q}, \varnothing, \varnothing)$, which can be interpreted as follows:

\begin{equation}
\label{eq:cfg}
\begin{split}
    \tilde{\bm{\epsilon}}_\theta(\bm{z}_t, t, \bm{z}_{q}, \bm{a}_e^1, \bm{a}_e^2)) = & \ \lambda \cdot \bm{\epsilon}_\theta(\bm{z}_t, t, \bm{z}_{q}, \bm{a}_e^1, \bm{a}_e^2) \\
    & + (1-\lambda) \cdot \bm{\epsilon}_\theta(\bm{z}_t, t, \bm{z}_{q}, \varnothing, \varnothing)
\end{split}
\end{equation}

\mypar{Recovering waveform.}
After obtaining the predicted noise $\Tilde{\epsilon}_\theta$ from the diffusion model, we first retrieve the encoded latent of the transformed mel-spectrogram. This latent is then decoded through the VAE decoder, and finally, the waveform is recovered using a pre-trained HiFi-GAN vocoder \cite{kong2020hifi}.

\section{Experiments}
\subsection{Experimental Setup}
\mypar{Dataset.}
Our goal is to train and evaluate the model's capability add, remove, and replace environmental sounds from an audio. For this, we utilize a combination of speech and environmental sound datasets:
\begin{itemize}[topsep=0pt, noitemsep, leftmargin=*]
\item \textbf{VCTK} \cite{yamagishi2019vctk}: The VCTK corpus contains 50 hours of speech data from 109 English speakers with various accents. We leverage this dataset as the source of clean speech examples for our training data.
\item \textbf{LibriSpeech} \cite{panayotov2015librispeech}: The LibriSpeech corpus contains around 1000 hours of English speech derived from audiobooks. We use a 100-hour subset of this corpus as an additional source of speech data for training and evaluation.
\item \textbf{BBC SFX} \cite{bbc2017}: The BBC SFX dataset comprises 40 hours of ambient sounds from the BBC News. It is used to provide the environmental sound textures for our manipulation tasks. In addition, it provides explicit annotations of audio types, allowing both methods to operate under the same conditions.
\end{itemize}

\begin{figure*}[t]
\centering  
\includegraphics[width=\textwidth]{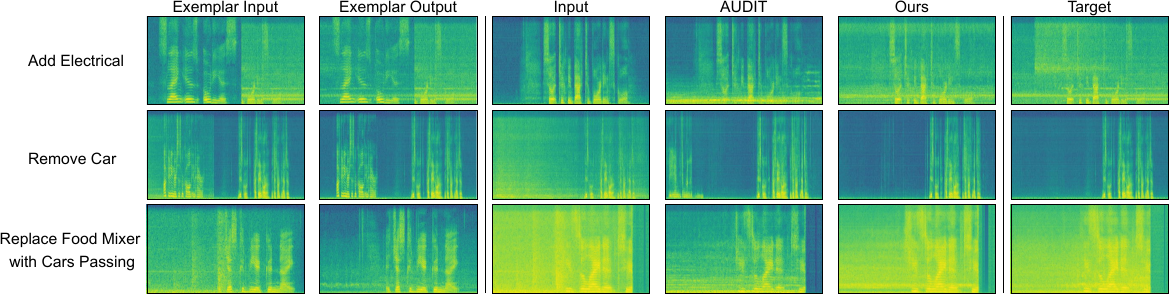}  
\caption{{\bf Model comparison.} We present qualitative results between our model and AUDIT, where each input audio is transformed according to the exemplar pairs.}  
\label{fig:crop_spectrograms}  
\vspace{-2mm}
\end{figure*}

\begin{figure}[t]
\centering  
\includegraphics[width=\linewidth]{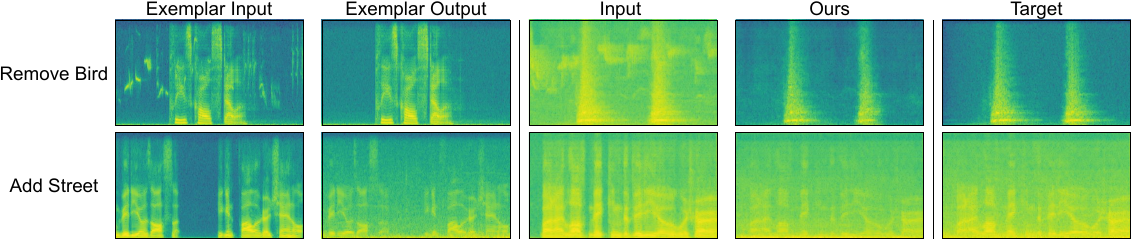}  
\caption{{\bf Generalization to real-world data.} Our model can generalize to non-speech (top) and real-world (bottom) scenarios.}  
\label{fig:real_world}  
\vspace{-4mm}
\end{figure}

\mypar{Model configurations.}
We use the pre-trained VAE encoder and decoder, and HiFi-GAN vocoder from AudioLDM\cite{liu2023audioldm}, which are trained on AudioCaps\cite{kim2019audiocaps}, AudioSet\cite{gemmeke2017audio}, BBC Sound Effect\cite{bbc2017}, and Freesound\cite{fonseca2021fsd50k} datasets. The VAE is configured with a compression level of 4 and a latent channels of 8. To extract the audio embedding from the exemplar pair, we fine-tune the CLAP audio encoder\cite{elizalde2023clap} along with the diffusion process.
The latent diffusion model features a U-Net-based backbone\cite{ronneberger2015u}, consisting of four encoder blocks for downsampling, a middle block, and four decoder blocks for upsampling. Multi-head attention\cite{vaswani2017attention} is applied in the last three encoder blocks and the first three decoder blocks. We employ $N=1000$ steps and a linear noise schedule ranging from $\beta_1=0.0015$ to $\beta_N=0.0195$ in the forward process. Moreover, we utilize the DDIM sampler \cite{song2020denoising} with 200 sampling steps for generating the output. We set the guidance scale $\lambda=4.5$ as described in Equation~\eqref{eq:cfg}.

\mypar{Training procedures.}
To ensure training efficiency, we first divide all audio files from the BBC SFX dataset into 10-second segments. We then split the VCTK, LibriSpeech, and BBC datasets into training and testing sets, with 150 hours of audio for training and 10 hours for testing. For each training step, we randomly select two 2.56-second speech samples from either the VCTK or LibriSpeech. One sample is used as the speech in the exemplar pair (pre-editing), and the other is used as the speech for the input to be transformed. To simulate the addition, removal, and replacement operations, ambient sounds from the BBC SFX dataset are randomly selected and added to the speech samples on the fly during training. We train our model over 100 epochs using the AdamW optimizer~\cite{loshchilov2017decoupled} with a learning rate of $10^{-4}$, $\beta_1=0.95$, $\beta_2=0.999$, $\epsilon=10^{-6}$, and a weight decay of $10^{-3}$.

\mypar{Evaluation metrics.}
To evaluate our model performance, we include objective metrics, including Fréchet Audio Distance (FAD) \cite{kilgour2018fr}, Fréchet Distance (FD) \cite{liu2023audioldm}, Kullback-Leibler Divergence (KL) \cite{kullback1951information},  Log Spectral Distance (LSD), Inception Score (IS), Perceptual Evaluation of Speech Quality (PESQ) \cite{rix2001perceptual}, Short-Time Objective Intelligibility (STOI) \cite{taal2010short}, Overall Quality (OVL), and Relevance to Intended Transformation (REL). FAD and FD both indicate the similarity between target and transformed audios, while KL quantifies how target and transformed audios differ in distribution. LSD measures the difference between frequency spectrograms of the target and transformed audios. IS and PESQ assess the diversity and quality of transformed audio. STOI measures how much a transformed speech signal is understandable to a listener. OVL represents the overall audio quality, while REL indicates the relevance between exemplar output and transformed audio. Both scores are rated on a scale from 1 to 5, where higher rating denotes better performance.

\noindent{\bf Baselines.} We compare our approach against the following baseline models:
\begin{itemize}[topsep=0pt, noitemsep, leftmargin=*]
\item \textbf{AUDIT} \cite{wang2023audit}: AUDIT is a text-conditioned diffusion model for audio editing based on user-provided text prompts. The template is defined as ``add source A," ``remove source A," and ``replace source A with source B." Since this model is not open source, we re-implement it and use it as the baseline for comparison.
\item \textbf{MP-SENet} \cite{lu2023mp}: MP-SENet is a GAN-based state-of-the-art speech enhancement model that performs parallel denoising of both magnitude and phase spectrograms. We include this model in the speech enhancement benchmark to evaluate our method's generalization capability.
\end{itemize}

\begin{table*}[t]
\centering
\caption{Quantitative objective results for the audio texture manipulation task, where the numbers (1, 2, 3) between the ``$\rightarrow$" notation represent the number of ambient sounds in the audio.} 
\label{tab:comb_quan}

\centering
{\scriptsize (a) Addition task} 
\label{tab:addition}
\vspace{0.1cm}
\resizebox{1.0\textwidth}{!}{
    \begin{tabular}{l|ccccc|ccccc|ccccc}
    \toprule
    & \multicolumn{5}{|c|}{Addition 0$\rightarrow$1} & \multicolumn{5}{c|}{Addition 1$\rightarrow$2} & \multicolumn{5}{c}{Addition 2$\rightarrow$3}\\
    \midrule
    \textbf{Method}  & \textbf{FAD} $(\downarrow)$ & \textbf{FD} $(\downarrow)$ & \textbf{KL} $(\downarrow)$ & \textbf{LSD} $(\downarrow)$ & \textbf{IS} $(\uparrow)$ &  \textbf{FAD} $(\downarrow)$ & \textbf{FD} $(\downarrow)$ & \textbf{KL} $(\downarrow)$ & \textbf{LSD} $(\downarrow)$  & \textbf{IS} $(\uparrow)$ & \textbf{FAD} $(\downarrow)$ & \textbf{FD} $(\downarrow)$ & \textbf{KL} $(\downarrow)$ & \textbf{LSD} $(\downarrow)$ & \textbf{IS} $(\uparrow)$ \\
    \midrule
    AUDIT  \cite{wang2023audit}      &4.77   &16.00   & 1.10   & 1.57   & \textbf{1.70}   & 3.44   & 11.43   & \textbf{0.78}   & 1.32  & \textbf{1.76}  & 3.46  & 11.13  & \textbf{0.81}  & 1.30  & 1.71  \\
    \midrule
    Ours         & 5.58    & 19.53 & 1.12 & 1.48 & 1.59 & 3.78  &12.62  &   0.92  & 1.37  & 1.68  & 3.92 & 11.47    & 0.89    & 1.31  & \textbf{1.75} \\
    + PE     & \textbf{3.83}    & \textbf{15.54}  & \textbf{0.99}  &  \textbf{1.40} & 1.60 & \textbf{2.68} &   \textbf{10.43} & 0.82 & \textbf{1.30} & 1.71 & \textbf{2.91}  & \textbf{10.93} & 0.88 & \textbf{1.26}    & 1.60    \\
    \bottomrule
    \end{tabular}
}

\vspace{0.3cm}

\centering
{\scriptsize (b) Removal task}
\label{tab:removal}
\vspace{0.1cm}
\resizebox{1.0\textwidth}{!}{
    \begin{tabular}{l|ccccc|ccccc|ccccc}
    \toprule
    & \multicolumn{5}{|c|}{Removal 1$\rightarrow$0} & \multicolumn{5}{c|}{Removal 2$\rightarrow$1} & \multicolumn{5}{c}{Removal 3$\rightarrow$2} \\
    \midrule
    \textbf{Method}  & \textbf{FAD} $(\downarrow)$ & \textbf{FD} $(\downarrow)$ & \textbf{KL} $(\downarrow)$ & \textbf{LSD} $(\downarrow)$ & \textbf{IS} $(\uparrow)$ &  \textbf{FAD} $(\downarrow)$ & \textbf{FD} $(\downarrow)$ & \textbf{KL} $(\downarrow)$ & \textbf{LSD} $(\downarrow)$ & \textbf{IS} $(\uparrow)$ & \textbf{FAD} $(\downarrow)$ & \textbf{FD} $(\downarrow)$ & \textbf{KL} $(\downarrow)$ & \textbf{LSD} $(\downarrow)$ & \textbf{IS} $(\uparrow)$  \\
    \midrule
    AUDIT \cite{wang2023audit}        &5.95   &9.20   & 0.38   & 1.75   & \textbf{1.36}   & 4.80   & 13.30   & \textbf{1.07}   &  1.80 & \textbf{1.72}  & 6.26  & 18.47  & 1.09  & 1.74  & \textbf{1.64}  \\
    \midrule
    Ours         & 5.86    & 7.51 & 0.29 & \textbf{1.61}    & 1.29 & 4.80  &13.31  &   \textbf{1.07}  & \textbf{1.63} & \textbf{1.72}    & 4.82 & 16.70    & \textbf{1.01}    &  \textbf{1.45}   & 1.56 \\
    + PE     &  \textbf{4.78}   & \textbf{5.87}    &  \textbf{0.26}   & 1.69  &  1.27   & \textbf{3.69} & \textbf{12.64}    & 1.21    & 1.67 & 1.66    & \textbf{3.67} & \textbf{13.11}  &  1.03 & 1.50 & 1.58 \\
    \bottomrule
    \end{tabular}
}

\vspace{0.3cm}

\centering
{\scriptsize (c) Replacement task}
\label{tab:replacement}
\vspace{0.1cm}
\resizebox{1.0\textwidth}{!}{
    \begin{tabular}{l|ccccc|ccccc|ccccc}
    \toprule
    & \multicolumn{5}{|c|}{Replacement 1$\rightarrow$1} & \multicolumn{5}{c|}{Replacement 2$\rightarrow$2} & \multicolumn{5}{c}{Replacement 3$\rightarrow$3} \\
    \midrule
    \textbf{Method}  & \textbf{FAD} $(\downarrow)$ & \textbf{FD} $(\downarrow)$ & \textbf{KL} $(\downarrow)$ & \textbf{LSD} $(\downarrow)$ & \textbf{IS} $(\uparrow)$  &  \textbf{FAD} $(\downarrow)$ & \textbf{FD} $(\downarrow)$ & \textbf{KL} $(\downarrow)$ & \textbf{LSD} $(\downarrow)$ & \textbf{IS} $(\uparrow)$ & \textbf{FAD} $(\downarrow)$ & \textbf{FD} $(\downarrow)$ & \textbf{KL} $(\downarrow)$ & \textbf{LSD} $(\downarrow)$ & \textbf{IS} $(\uparrow)$  \\
    \midrule
    AUDIT \cite{wang2023audit}       & \textbf{4.94}   &\textbf{15.86}   & 1.28   & 1.62   & 1.57   & \textbf{4.31}   & 15.62   & 1.20   & 1.55  & 1.54  & \textbf{3.12}  & 16.35  & 1.06  & 1.48  & 1.47  \\
    \midrule
    Ours         & 6.58    & 18.34 & 1.13 & 1.59 & 1.63 & 4.70  &14.31  &   \textbf{1.18}  & 1.51    & 1.63    & 3.61 & 13.02    & 0.93    & 1.41  & 1.52 \\
    + PE     & 5.80    & 17.21  & \textbf{1.09} & \textbf{1.58} & \textbf{1.68}    & 4.40    & \textbf{13.48}    & 1.19    & \textbf{1.49}  &  \textbf{1.67}   & 4.36  & \textbf{11.53} & \textbf{0.90}     & \textbf{1.34} &  \textbf{1.57}   \\
    \bottomrule
    \end{tabular}
}
\vspace{-2.5mm}
\end{table*}

\begin{table}[t]
\scriptsize
\centering
\caption{Quantitative generalization results on speech enhancement.}
\begin{tabular}{lcc}
\toprule
\textbf{Method}     & \textbf{PESQ} $(\uparrow)$ & \textbf{STOI} $(\uparrow)$ \\
\midrule
Noisy      &   1.72        &     0.79       \\
\midrule
MP-SENet~\cite{lu2023mp}   &   2.55        &     0.91      \\
\midrule
Ours       &   1.82        &     0.85       \\
\bottomrule
\end{tabular}
\label{tab:se}
\vspace{-3mm}
\end{table}

\subsection{Comparison to Baselines}
\mypar{Quantitative results.}
We start by presenting the quantitative results for the addition, removal, and replacement tasks in Tab. \ref{tab:addition} (a), \ref{tab:removal} (b), and \ref{tab:replacement} (c). For the addition task, we evaluate three cases: adding noise to speech with zero, one, or two background noises. For the removal task, we assess the model's ability to remove noise from speech with one, two, and three background noises. In the replacement task, we evaluate replacing one noise with another in speech that contains one, two, or three noises. The results show that our model consistently outperforms AUDIT in both the addition and removal tasks across all metrics. Although our model does not surpass AUDIT in the $1\rightarrow1$ replacement case, it performs better when more background noises are present, demonstrating stronger capability in more complex replacement tasks.

To further validate our model's performance, we conduct a human evaluation, as shown in Tab. \ref{tab: quan_sub}. We randomly select 20 generated audio samples from the test set, rated by 30 participants, with the average and 95\% confidence intervals calculated for each metric as the final results. To ensure the reliability of the evaluation, we include two control sets --- one with pure noise and the other with ground-truth audio. The participants consistently prefer audio generated by our model, which aligns with the objective evaluation results.

\mypar{Qualitative results.}
We also visualize the results of our model across various tasks using exemplar pairs and compare them with baselines and ground truth in Fig. \ref{fig:crop_spectrograms}. In the addition task, our model generates audio textures that match the exemplar pair, whereas AUDIT fails to capture high-frequency details. For the removal task, AUDIT struggles to remove a loud car sound, while our model captures the texture changes from the exemplar pair, even when the sound is less pronounced in the exemplar. In the replacement task, AUDIT suppresses high-frequency regions, while our model smoothly reduces their intensity, effectively resembling the transformation specified by the exemplar pair. These results demonstrate our model's superior ability to replicate sound texture transformations guided by the exemplar pair.

\subsection{Ablation Study and Analysis}
We conduct an ablation study to evaluate our model with and without learnable positional encoding in Tab. \ref{tab:comb_quan}, which is designed to capture the temporal relationship between pre-editing and post-editing audio clips. The results show that incorporating positional encoding improves performance, highlighting its importance in understanding transformation direction. 

\subsection{Evaluation on Speech Enhancement}
\label{subsec: eval_se}
Although our model is not specifically trained for speech enhancement. we evaluate its ability to generalize to this task. Unlike the removal task, where the model focuses on removing a particular sound texture based on the exemplar pair, speech enhancement requires removing all background noise from the speech. We compare our model with MP-SENet \cite{lu2023mp} on the VoiceBank + DEMAND dataset \cite{valentini2016investigating}. Please note that the evaluation data is processed through the HiFi-GAN vocoder for consistency across models. Since PSEQ is a phase-aware metric, the vocoder-reconstructed audio does not retain phase information, which results in lower PSEQ scores.

Despite these challenges, we demonstrate that our model can enhance speech and generalize to out-of-distribution data. However, because the exemplar pairs are randomly sampled, there are instances where the exemplar input's noise texture does not closely match that of the noisy input. This mismatch can confuse the model and hinder its ability to apply the correct transformation. As a result, our model does not perform as well as MP-SENet on this specific task (Tab.~\ref{tab:se}).

\subsection{Generalization to Real-World Data}
We ask whether our model can generalize to non-speech audios and ``in-the-wild" datasets. As shown in Fig. \ref{fig:real_world}, our model can remove the bird sound from non-speech inputs, such as dog barking. In addition, we test our model on the {\em CityWalk} dataset \cite{li2024self}, demonstrating its ability to generalize to real-life scenarios.

\begin{table}[t]
    \scriptsize
    \centering
    \caption{Quantitative subjective results on the CityWalk dataset, where OVL and REL are presented with 95\% confidence intervals.}
    \label{tab: quan_sub}
    \begin{tabular}{lcc}
        \toprule
        \textbf{Method} & \textbf{OVL} $(\uparrow)$ & \textbf{REL} $(\uparrow)$\\
        \midrule
        GT & $4.21 \pm 0.07$ & $4.15 \pm 0.05$ \\
        \midrule
        AUDIT \cite{wang2023audit} & $3.12 \pm 0.06$ & $3.26 \pm 0.04$ \\
        \midrule
        Ours & $\textbf{3.76} \boldsymbol{\pm} \textbf{0.05}$ & $\textbf{3.88} \boldsymbol{\pm} \textbf{0.05}$ \\
        \bottomrule
    \end{tabular}
    \vspace{-0.15mm}
\end{table}

\vspace{2.2mm}
\section{Conclusion}
In this paper, we introduced an exemplar-based analogy model for audio texture manipulation, aiming to learn transformations from paired speech examples. By constructing a quadruplet dataset of speech and ambient sounds, we trained a latent diffusion model to perform tasks such as adding, removing, and replacing specific auditory elements. Objective metrics and perceptual studies demonstrate the effectiveness of our model in manipulating audio textures, complementing text-conditioned baselines, and generalizing well to real-world, out-of-distribution, and non-speech scenarios. We hope this work opens new possibilities for more flexible audio editing methods and provides a foundation for future research on exemplar-based transformations in speech and broader audio domains.

\mypar{Limitations.}
Our model shows promising results on several tasks but has some limitations. It needs paired data for training, so it cannot use the abundantly available internet data, which restricts the model's ability to learn from diverse real-world audio textures. Also, as noted in Sec.~\ref{subsec: eval_se}, it may produce poor results if the transformation direction is not clearly specified. Finally, our method does not support precise position editing for the add operation, as it focuses on global texture modifications rather than spatial placement.

\mypar{Acknowledgment.} We thank Yisi Liu for the helpful discussion. This work was funded by the Society of Hellman Fellows.

\clearpage
\bibliographystyle{IEEEtran}
\bibliography{main}
\end{document}